\documentclass[
aps,%
12pt,%
final,%
notitlepage,%
oneside,%
onecolumn,%
nobibnotes,%
nofootinbib,%
superscriptaddress,%
showpacs,%
centertags]%
{revtex4}
\usepackage{showkeys}
\usepackage{graphicx}
\def\DR{\rm I\kern-1.45pt\rm R}
\def\DC{\kern2pt {\hbox{\sqi I}}\kern-4.2pt\rm C}

\newcommand{\ba}{\begin{array}}
\newcommand{\ea}{\end{array}}
\newcommand{\be}{\begin{equation}}
\newcommand{\ee}{\end{equation}}
\newcommand{\bea}{\begin{eqnarray}}
\newcommand{\eea}{\end{eqnarray}}
\newcommand{\bi}{\begin{itemize}}
\newcommand{\ei}{\end{itemize}}

\usepackage{amscd,amsmath,amssymb}

\begin{document}
\title{Phase diagrams of the Ising--Heisenberg chain with S=1/2 triangular  $XXZ$ clusters.}
\author{\firstname{Vadim}~\surname{Ohanyan}}
\affiliation{Yerevan State University, Yerevan, Armenia}
\begin{abstract}
The one dimensional spin system consisted of triangular $S=1/2$ $XXZ$ Heisenberg clusters
alternating with single Ising spins is considered. Partition function of the system is
calculated exactly within the transfer--matrix formalism. $T=0$ ground state phase diagrams,
corresponding to different regions of the values of system parameters are obtained.
\end{abstract}

  \maketitle
\section{Introduction}
    The Heisenberg model, on the one hand, is the key component of
 microscopic physics underlying the magnetic properties of
 materials \cite{1}, on the other hand, its one-dimensional variant plays one of the
 fundamental roles in mathematical physics as the example of integrable quantum many particle
 systems \cite{Bax}. However, applicability of its exact
 solutions (which exists only in one-dimensional case)
 in describing the thermodynamic properties of real magnetic materials are rather
 limited, because of fact that in given context the term "exact solution"
means rather possibility to obtain quantum-mechanical spectrum of
the system in principle, than the possibility of obtaining analytic
expressions for thermodynamic functions. So, the Bethe ansatz
technique, which is applicable in simplest case of Heisenberg spin
chains, leads to the transcendent equation for determining the
spectral parameters \cite{bethe}, but it is almost useless in
calculating the partition function. Though, there exists rather
sophisticated analytic technique for calculating the finite-T
properties of Heisenberg spin chains \cite{klu}, it is still of
current importance to find alternative relatively simple approximate
methods of describing the thermodynamic properties of the model of
real magnetic materials. One of such methods consists in the
changing of some (or even all) exchange Heisenberg interactions
between the spin with more simple "classical" ones, which allows one
to develop the exact classical transfer-matrix technique for
obtaining analytic expression for partition function and all
thermodynamic functions. This approach has been shown to be in
satisfactory qualitative(even quantitative in some cases) agreement
with the experimental data and numerical calculations for some
variant of one-dimensional spin chains, namely in F-F-A and F-F-A-A
alternating chains \cite{oha}-\cite{ant}. In this work we consider
the Ising-Heisenberg chain with triangular Heisenberg XXZ clusters
alternating with single Ising spins, the interaction between spins
in each triangle and its adjoint two Ising spin is assumed to be of
Ising type which allow one to implement the exact analytic
calculation of all thermodynamic properties.

 \begin{figure}
 \begin{center}
  \includegraphics{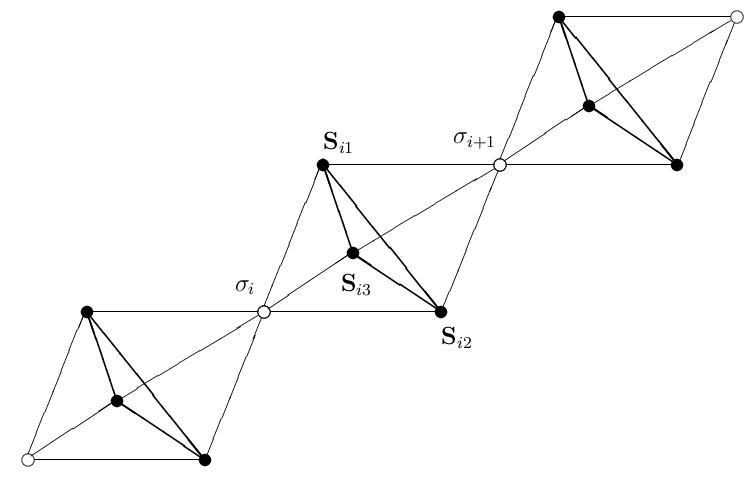}
  \caption{The schematic picture of the Ising--Heisenberg chain with triangular plaquettes alternating with single sites.
  The filled(empty) circles indicate Heisenberg(Ising) spins.\label{fig1}}
           \end{center}
         \end{figure}

 \section{Solution of the general model of Heisenberg-Ising chain}
 Let us consider the system of Heisenberg and Ising spins residing at the one-dimensional chain in the following way.
 Heisenberg spins form clusters with relatively small number of sites, within these cluster the spins are interact
 with exchange interaction, another kinds of interaction, e.g. Dzyaloshinskii-Moriya term, single-ion anisotropy,
 dipole-dipole interaction, Zeeman terms, e.t.c. can be presented as well. The topology of links between the Heisenberg
 spins can be various, in most simple cases it could be finite linear chain or close loop.  The Heisenberg spin clusters
 compose the chain by alternating with the single sites with Ising spins. This means that quantum spins from different cluster do not
 interact to each other, but only with the Ising spins between them. Interaction between Heisenberg and Ising spins is supposed
 to be of Ising type, i.e. $S^z \sigma$. The Hamiltonian for such a model can be written in the following way
 \bea
 \mathcal{H}=\sum_{i=1}^N\left( \mathcal{H}_i \left(\mathbf{S}_{i1},...,\mathbf{S}_{im} \right)-\frac{H_2}{2}\left(\sigma_i+\sigma_{i+1} \right)\right), \label{ham}
 \eea
 where $\mathcal{H}_i \left(\mathbf{S}_{i1},...,\mathbf{S}_{im} \right)$ is the Hamiltonian of the $i-th$ Heisenberg cluster, containing $m$ sites, $\sigma_i$ and $\sigma_{i+1}$ are two its adjacent Ising spins, takin value $\pm 1$ and $H_2$ stand for the dimensionless magnetic field acting on Ising spins. One can assume for a while that the corresponding field acting on Heisenberg spins via $-H_1 S_{ia}^z$ differs from the $H_1$ which physically can be explained by the difference in the $g$-factors. The term describing interaction between the Heisenberg spins of $i$-th cluster and $\sigma_i$ and $\sigma_{i+1}$ is included into $\mathcal{H}_i$ and in general case has the form:
 \bea
 \mathcal{H}_i^{S\sigma}= \sum_{a=1}^m \left( \sigma_i L_a +\sigma_{i+1} R_a\right)S_{i,a}^z, \label{hamSs}
 \eea
 where $L_a(R_a)$, $a=1,...,m$ is the set of the coupling constants of the interaction between $a$-th spin from the cluster and its left(right) adjacent Ising spin. The absence of the immediate exchange interaction between spins from different clusters leads to commutativity of the Hamiltonians for different clusters
 \bea
 [\mathcal{H}_i,\mathcal{H}_j]=0. \label{comu}
 \eea
 This fact makes possible to use the classical transfer-matrix technique for obtaining the exact thermodynamic solution for such kind of spin chains. Namely, consider the partition function
  \bea
 {\mathcal{Z}}=\sum_{(\sigma)}\mbox{Tr}_{(S)}e^{-\beta\sum_{i=1}^N \left({\mathcal{H}}_i-\frac{H_2}{2}\left(\sigma_i+\sigma_{i+1} \right)\right)},
 \label{z1}
 \eea
  where the sum is going over all possible configurations of the
 Ising spins and $\mbox{Tr}_{(S)}$ denotes the trace over all
 Heisenberg operators $S$ and $\beta$ as usually is the inverse temperature. Commutativity of the Hamiltonians for different cluster allows one
 to expand the exponent in the partition function:
 \bea
&&{\mathcal{Z}}=\sum_{(\sigma)}\prod_{i=1}^N e^{\beta\frac{H_2}{2}\left(\sigma_i+\sigma_{i+1} \right)} Z(\sigma_i, \sigma_{i+1})=\sum_{(\sigma)}\prod_{i=1}^N T(\sigma_i, \sigma_{i+1}), =\mbox{Tr}\mathbf{T}^N\\ \nonumber\label{z2}
 &&Z(\sigma_i, \sigma_{i+1})=\mbox{Tr}_i
e^{-\beta{\mathcal{H}}_i}=\sum_{n=1}^D\langle
\psi_n|e^{-\beta {\mathcal{H}}_i}|\psi_n\rangle=\sum_{n=1}^D e^{-\beta \lambda_n(\sigma_i,
\sigma_{i+1})},
 \eea
 here $\mbox{Tr}_i$ stands for the trace over the state of i-th
 Heisenberg spin cluster and $D=\prod_{k=1}^m(2s_k+1)$ is the dimensionality of the corresponding matrix, $s_k$ is the spin of k-th spin in the cluster. Thus, the partition function takes the form of the partition function of one-dimensional chain with classical discrete two-state variable on each site.
The corresponding transfer-matrix ${\mathbf{T}}$ reads : \bea
{\mathbf{T}}=\left( \begin{array}{lcr}
      e^{\beta H_2}Z_+  & Z_0 \\
      \bar{Z_0}  & e^{-\beta H_2}Z_- \label{T}
      \end{array}
\right), \eea
where $Z_+=Z(1,1), Z_-=Z(-1,-1), Z_0=Z(1,-1)$ and $\bar{Z_0}=Z(-1,1)$.
According to Eq. (\ref{z2}), the partition function is the sum of N-th powers of the eigenvalues of  ${\mathbf{T}}$, which are given by: \bea
\Lambda_{\pm}=\frac{1}{2}\left(e^{\beta H_2}Z_+ +e^{-\beta H_2}Z_-
\pm \sqrt{\left(e^{\beta H_2}Z_+ -e^{-\beta H_2}Z_-
\right)^2+4Z_0\bar{Z_0}} \right). \label{lamb}
 \eea
 Thus,
the free energy per one spin in the thermodynamic limit when only
maximal eigenvalue survive is \bea f=-\frac{1}{(m+1) \beta}\log
\left(\frac{1}{2}\left(e^{\beta H_2}Z_+ +e^{-\beta H_2}Z_- +
\sqrt{\left(e^{\beta H_2}Z_+ -e^{-\beta H_2}Z_- \right)^2+4Z_0\bar{Z_0}}
\right) \right), \label{fe_t}
 \eea
 The different variant of Ising--Heisenberg chains with two- and three-spin Heisenberg
 clusters with the properties described above are considered in Refs.
 \cite{str}-\cite{ant}. Particularly, in Ref. \cite{ant} the Ising--Heisenberg chain with triangular XXZ spin-1/2 clusters was solved exactly within the technique described above. The system is described by Hamiltonian given by Eq.(\ref{ham}) with
 \bea
 &&{\mathcal{H}}_i=J\left(\frac{\Delta}{2}(S_{i1}^+ S_{i2}^-+S_{i1}^-
S_{i2}^+)+S_{i1}^zS_{i2}^z+\frac{\Delta}{2}(S_{i1}^+
S_{i3}^-+S_{i1}^-
S_{i3}^+)+S_{i1}^zS_{i3}^z+\frac{\Delta}{2}(S_{i2}^+
S_{i3}^-+S_{i2}^- S_{i3}^+)+S_{i2}^zS_{i3}^z\right)\\ \nonumber
&&+K(S_{i1}^z+S_{i2}^z+S_{i3}^z)(\sigma_i+\sigma_{i+1})-H_2(S_{i1}^z+S_{i2}^z+S_{i3}^z),
 \eea
 All free spins from each triangle interact with their two adjacent Ising spin with the same constant $K$ (See Fig. (\ref{fig1})).
 To construct the transfer matrix $\mathbf{T}$ one should obtain eight eigenvalues $\lambda_n(\sigma_i, \sigma_{i+1} )$ of these Hamiltonian, which are the following
 \bea
&& \lambda_{1,2}(\sigma_i, \sigma_{i+1})=3(J\pm H)\mp
 3K(\sigma_i+\sigma_{i+1}), \label{eig} \\ \nonumber
 &&\lambda_{3,4}(\sigma_i,
 \sigma_{i+1})=\lambda_{5,6}(\sigma_i, \sigma_{i+1})=-(1+2\Delta)J \pm H \mp
 K(\sigma_i+\sigma_{i+1}), \\ \nonumber
 &&\lambda_{7,8}(\sigma_i, \sigma_{i+1})=-(1-4\Delta)J \pm H \mp
 K(\sigma_i+\sigma_{i+1}). \label{ev}
 \eea
 Thus
 \bea
&&Z(\sigma_i, \sigma_{i+1})=B_1
\cosh(\beta(H_2-K(\sigma_i+\sigma_{i+1})))+B_2\cosh(\beta3(H_2-K(\sigma_i+\sigma_{i+1}))),\\ \nonumber
\label{zss}
&& B_1=2\left(e^{\beta(1-4 \Delta) J}+2e^{\beta(1+2 \Delta) J}
 \right), \quad B_2=2e^{-\beta 3 J}. \label{BB}
 \eea
 Having all these functions one can easily obtain the partition function and free energy and then, exploiting general thermodynamic relations, taking a derivatives of it, various thermodynamic functions, like magnetization, specific head, susceptibility, e.t.c. It Ref. (\cite{ant}) these issue was presented in details for the case of antiferromagnetic coupling. Here we analyze the ground state properties of the system in all range of coupling constants.

\section{T=0 Phase diagrams}

\begin{figure}
 \begin{center}
  \includegraphics{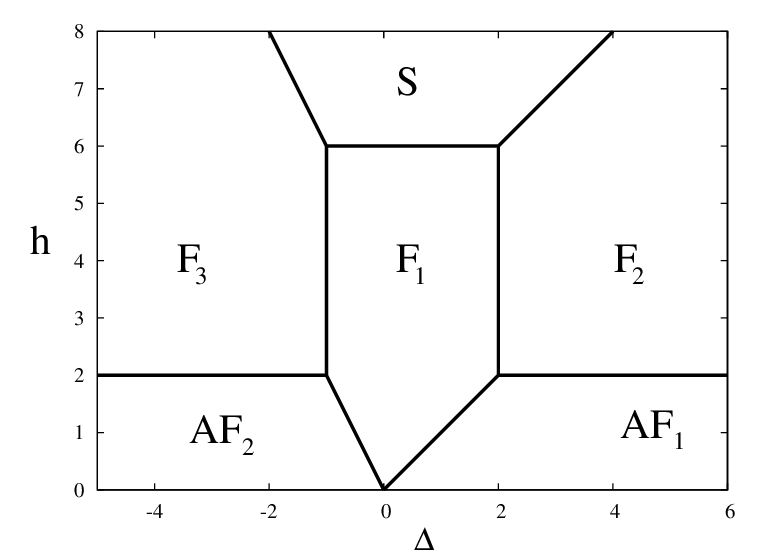}
  \caption{$T=0$ ground state phase diagram for $J>0, K>0$ and $\eta=1$\label{fig2}}
           \end{center}
         \end{figure}

\begin{figure}
 \begin{center}
  \includegraphics{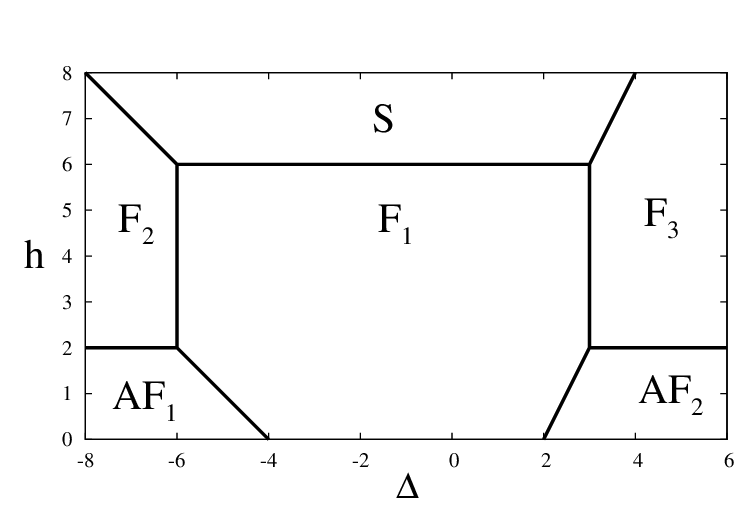}
  \caption{$T=0$ ground state phase diagram for $J<0, K>0$ and $\eta=1$.\label{fig3}}
           \end{center}
         \end{figure}

\begin{figure}
 \begin{center}
  \includegraphics{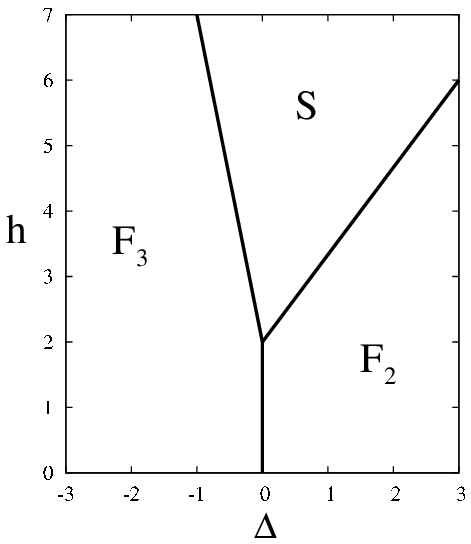}
  \caption{$T=0$ ground state phase diagram for $J>0, K<0$ and $\eta=2$\label{fig4}}
           \end{center}
         \end{figure}

\begin{figure}
 \begin{center}
  \includegraphics{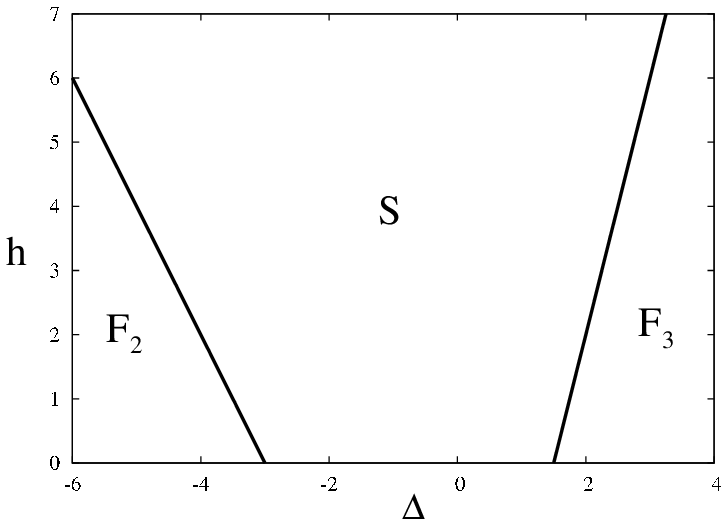}
  \caption{$T=0$ ground state phase diagram for $J<0, K<0$ and $\eta=2$\label{fig5}}
           \end{center}
         \end{figure}

 Let us write down the possible microscopic configurations of spins in the chain with corresponding
 energies per one lattice unit (in our case each block contains one triangle and its one adjoint Ising spin)
 and magnetizations. Up to the inversion of the direction of total magnetization, one can obtain six different
 eigenstates of the model under consideration: saturated completely polarized state (S),
 three ferrimagnetic states, with magnetization $M=1/2$, $(F_1), (F_2), (F_3)$ and two antiferromagnetic
 states with $M=0$, ($AF_1$) and ($AF_2$):
 \bea
 &&|S\rangle=\prod_{i=1}^N|3/2,3/2\rangle_i\bigotimes|\uparrow\rangle_i,\quad E_S=3J+6K-4H,\quad M=1, \\ \nonumber
 &&|F_1\rangle=\prod_{i=1}^N|3/2,3/2\rangle_i\bigotimes|\downarrow\rangle_i,\quad E_{F_1}=3J-6K-2H,\quad M=1/2, \\ \nonumber
 &&|F_2\rangle=\prod_{i=1}^N|1/2,1/2\rangle_i\bigotimes|\uparrow\rangle_i,\quad E_{F_2}=-J(1+2 \Delta)+2K-2H,\quad M=1/2, \\ \nonumber
 &&|F_3\rangle=\prod_{i=1}^N|3/2,1/2\rangle_i\bigotimes|\uparrow\rangle_i,\quad E_{F_3}=-J(1-4 \Delta)+2K-2H,\quad M=1/2, \\ \nonumber
 &&|AF_1\rangle=\prod_{i=1}^N|1/2,1/2\rangle_i\bigotimes|\downarrow\rangle_i,\quad E_{AF_1}=-J(1+2\Delta)-2K,\quad M=0, \\ \nonumber
 &&|AF_2\rangle=\prod_{i=1}^N|3/2,1/2\rangle_i\bigotimes|\downarrow\rangle_i,\quad E_{AF_2}=-J(1-4\Delta)-2K,\quad M=0,  \label{st}
 \eea
 where $|l,m\rangle$ stand for the standard symmetry adopted states of three S=1/2 spins with  $S^{tot}=l$ and $S^{tot}_z=m$
 and $|\uparrow(\downarrow)\rangle$ are up(down) states of the corresponding Ising spins. Comparing the energies
 of different states one can obtain $T=0$ phase diagram of the model. In Ref. \cite{ant} the diagram was found only
 for antiferromagnetic values of all coupling constant, here we present the phase diagrams for all possible values of model parameters.
 In Fig. (\ref{fig2}) one can see the ground states phase diagram in the $(H-\Delta)$-plane for the case $J>0, K>0$ and particular
 value $\eta=|J/K|=1$. In this case all six ground states can be realized depending of the value of exchange anisotropy $\Delta$ and
 external magnetic field. For positive values of $\Delta$ one can see ($S$), ($F_1$), ($F_2$) and ($AF_1$)  states,
 whereas for $\Delta<0$  ($F_3$) and ($AF_2$) states appear instead of $(F_2)$ and $(AF_1)$ respectively.
 The boundary between $(AF_1)$ and $(F_1)$ states is given by $H=J(2+\Delta)-2K$, between $(AF_2)$ and $(F_1)$ by $H=2J(1+|\Delta|)-2K$,
 between $(S)$ and $(F_2)$ by $H=J(2+\Delta)+2K$ and finally between $(S)$ and $(F_3)$ by $H=2J(1+|\Delta|)+2K$.
 The ferrimagnetic phase $F_1$ passes to the saturated phase at $H=6K$, another horizontal line $H=6K$ in the diagram
 separates antiferromagnetic phases from ferrimagnetic ones. In case of $J<0, K>0$ (Fig. (\ref{fig3})) again one can see all six phases.
 The qualitative difference from previous case is the appearance of the wide gap between ($AF_1$) and ($AF_2$) states. It is also should
 be noted that the phases which were compatible with positive values of $\Delta$ in previous case now correspond to its negative values
 and vice versa. Like in the previous case two horizontal lines at $H=6K$ and $H=2K$ separate fully polarized phase from $(F_1)$ and
 antiferromagnetic phases from ferrimagnetic ones respectively. The boundaries between phases are defined by the following
 equations: $H=2|J|(\Delta-1)-2K$ between$(F_1)$ and$(AF_2)$; $H=|J|(|\Delta|-2)-2K$ between
 $(F_1)$ and $(AF_1)$;$|\Delta|=\frac{2(K+|J|)}{|J|}$ between $(F_1)$ and $(F_2)$; and $\Delta=\frac{K+|J|}{|J|}$ between
 $(F_1)$ and $(F_3)$. The next case, $J>0, K<0$ (Fig. (\ref{fig4})) is characterized by only three phases to appear, namely,
 $(F_2)$, $(F_3)$ and $(S)$. In this diagram two different ferrimagnetic phases are separated by $\Delta=0$ value. So, at
 negative values of $\Delta$ the ground state of the system is $(F_3)$, positive values of $\Delta$ lead to the $(F_2)$ ground state.
 Both ferrimagnetic phases pass immediately to saturated one at certain value of the magnetic field. For $(F_3)$ phase these value is
 given by $H=2J(1+|\Delta|)-2|K|$, for $(F_2)$ by $H=J(2+|\Delta|)-2|K|$. In Fig. (\ref{fig5}) the ground state phase diagram for
 $J<0, K<0$ is presented for $\eta=2$. Here again only three phases from previous case can be
 realized. For the values of axial anisotropy $\Delta$ belonging to
 the interval from $\Delta=-2\frac{|K|+|J|}{|J|}$ to
 $\Delta=\frac{|K|+|J|}{|J|}$, the only possible ground state of the
 system is $(S)$. For $\Delta <-2\frac{|K|+|J|}{|J|}$ the ground
 state became $(F_2)$ which pass to saturated one at
 $H=|J|(|\Delta|-2)-2|K|$. The same picture takes place for $\Delta
 >\frac{|K|+|J|}{|J|}$ with respect to the $(F_3)$ phase.

\section{Acknowledgements}
 Author would like to thank A. Badasyan for help in preparing the figures. This work was partly
supported by the grants CRDF-UCEP-06/07, ANSEF-1386-PS and
INTAS-05-7928.

\newpage

\newpage

\newpage

 \end{document}